\documentclass[aps,prb,twocolumn]{revtex4}
\usepackage{amsmath}
\usepackage{amsfonts, amsthm, amssymb}

\input{tcilatex}

\begin{document}

\title{Generalized DC and AC Josephson effects in antiferromagnets and in
antiferromagnetic $d$-wave superconductors}
\author{D. Chass\'{e}$^{1}$ and A.-M.S. Tremblay$^{1,2}$ }
\affiliation{1. D\'{e}partment de physique and RQMP, Universit\'{e} de Sherbrooke,
Sherbrooke, Qu\'{e}bec, Canada J1K 2R1.}
\affiliation{2. Canadian Institute for Advanced Research, Toronto, Ontario, Canada}
\date{\today }

\begin{abstract}
The Josephson effect is generally described as Cooper pair tunneling, but it
can also be understood in a more general context. The DC Josephson effect is
the pseudo-Goldstone boson of two coupled systems with a broken continuous
abelian $U(1)$ symmetry. Hence, an analog should exist for systems with
broken continuous non-Abelian symmetries. To exhibit the generality of the
phenomenon and make predictions from a realistic model, we study tunneling
between antiferromagnets and also between antiferromagnetic d-wave
superconductors. Performing a calculation analogous to that of Ambegaokar
and Baratoff for the Josephson junction, we find an equilibrium current of
the staggered magnetization through the junction that, in antiferromagnets,
is proportional to $\mathbf{\hat{s}}_{L}\times \mathbf{\hat{s}}_{R}$ where $%
\mathbf{\hat{s}}_{L}$ and $\mathbf{\hat{s}}_{R}$ are the N\'{e}el vectors on
either sides of the junction. Microscopically, this effect exists because of
the coherent tunneling of spin-one particle-hole pairs. In the presence of a
magnetic field which is different on either sides of the junction, we find
an analog of the AC Josephson effect where the angle between N\'{e}el
vectors depends on time. In the case of antiferromagnetic $d$-wave
superconductors we predict that there is a contribution to the critical
current that depends on the antiferromagnetic order and a contribution to
the spin-critical current that depends on superconducting order. The latter
contributions come from tunneling of the triplet Cooper pair that is
necessarily present in the ground state of an antiferromagnetic d-wave
superconductor. All these effects appear to leading order in the square of
the tunneling matrix elements.
\end{abstract}

\maketitle










\section{Introduction}

One of the most striking manifestations of superconductivity is the
Josephson effect, which consists in the coherent tunneling of Cooper pairs
across a junction when both sides are superconductors. Interestingly, this
occurs even though the Hamiltonian contains only single-electron tunneling. 
The resulting current is proportional to $\sin (\varphi )$, where $\varphi $
is the difference between the phases of the two superconducting order
parameters. In the presence of a constant potential difference, gauge
invariance requires the phase difference to increase linearly with time,
leading to an alternating current whose frequency depends only on universal
physical constants and not on any material parameter.

Superconductivity is just one example of coherence and spontaneous symmetry
breaking. The order parameter is a complex number so the broken symmetry is $%
U(1)$. The superconducting state selects a phase and the Josephson effect
arises from the tendency to make the phase uniform across the tunnel
junction.

Since there are many other types of order and corresponding broken
symmetries, the question of the analog of the Josephson effect in such cases
arises naturally \cite{Paranjape}. Indeed, one should expect differences in
order parameters across a junction to lead to a coherent tunneling of the
condensed objects that exist in the broken symmetry state. This question is
especially relevant in the context where junctions between magnetic
materials are of utmost importance for spintronics. As a matter of fact,
theoretical predictions have been made concerning the possible existence of
equilibrium spin currents \cite{Chandra:1990,Bruno:2005} of the Josephson
type in ferromagnetic (FM) tunnel junctions, in analogy with superconducting
junctions ~\cite{Lee, Nogueira, wang}. FM long range order is a realization
of spontaneous $SO(3)$ symmetry breaking and an equilibrium spin current
would result from the exchange coupling between the magnetic moments in the
two leads, which favors alignment of order parameters. From a
Ginzburg-Landau point of view, there is a term proportional to $\mathbf{M}%
_{L}\cdot \mathbf{M}_{R}$ where $\mathbf{M_{L}}$ and $\mathbf{M_{R}}$ are,
respectively, the magnetic moments on the left and on the right of the
junction. The Heisenberg equations of motion thus lead to $\tfrac{d\mathbf{%
M_{L}}}{dt}\sim \mathbf{M_{L}}\times \mathbf{M_{R}}$, corresponding to a
spin current.


In this paper, we study Josephson-like phenomena between antiferromagnets
(AF), normal and superconducting. Such phenomena are interesting for several
reasons. In the context of spintronics, it has been shown experimentally
that the absence of a net angular momentum in AF results in orders of
magnitude faster spin dynamics than in FM, which could expand the now
limited set of applications for AF materials \cite{Kimel}. More generally,
in AF the N\'{e}el order parameter breaks both lattice translation symmetry
and $SO(3)$ spin rotation symmetry so the situation is less straightforward
than in FM. In addition, AF are often close to superconducting phases, as is
found in heavy fermions, high-temperature superconductors and layered
organic superconductors. Understanding Josephson-like phenomena in AF is the
first step towards more general studies with coexisting antiferromagnetic
and superconducting order parameters, which we also pursue in this paper.
Generalized Josephson effects may help identify homogeneous coexistence of
antiferromagnetism and superconductivity in real materials. Studies along
these lines have recently appeared for ferromagnetism coexisting with $p-$%
wave superconductivity \cite{Sudbo:2007}.

We begin in Sec. II from a microscopic mean-field model for itinerant
antiferromagnets and for single electron tunneling and establish analogies
with the BCS ground state. This microscopic calculation leads in Sec. III to
an explicit expression for the analog of the critical current and its
temperature dependence. In addition, we show that (a) Cooper-pair tunneling
is replaced by tunneling of a spin-one neutral particle-hole pair (b) time
dependence introduced by external magnetic fields resemble the AC Josephson
effect but there are many differences because of the non-Abelian nature of
the problem and because spins do not couple to the gauge field but directly
to the magnetic field. We also discuss in Sec. IV how to observe this effect
by coupling to a ferromagnet through a tunnel junction. We then move in Sec.
V to the case of $d-$wave superconducting antiferromagnets and show that
there is a contribution to the Josephson charge current that is modulated by
the antiferromagnetic order parameter and, conversely, a contribution to the
spin-Josephson current that is modulated by the superconducting order
parameter. Our results are summarized in the discussion and conclusion
section. Additional details of the calculation may be found in Ref. \cite%
{Chasse:2009}. For higher-order effects in the tunneling matrix elements 
\cite{Anantram:2007,Shen:2007} that we do not discuss here, see also Ref. %
\onlinecite{Chasse:2009}.

\section{Model}

\subsection{Antiferromagnetic state and analogy with BCS ground state}

The Hamiltonian for a tunneling junction consisting of two leads of an AF
material and an insulating barrier between them reads 
\begin{equation}
H=H_{L}(c_{\mathbf{k}\sigma }^{\dag },c_{\mathbf{k}\sigma })+H_{R}(d_{%
\mathbf{q}\sigma }^{\dag },d_{\mathbf{q}\sigma })+H_{T}\;,
\end{equation}%
\noindent where $H_{L(R)}$ is the Hamiltonian of the left (right) AF, $H_{T}$
is the tunneling part connecting the two leads, and $c_{\mathbf{k}\sigma
}^{\dag }(c_{\mathbf{k}\sigma })$ and $d_{\mathbf{q}\sigma }^{\dag }(d_{%
\mathbf{q}\sigma })$ are the fermion creation (annihilation) operators of
the left and right leads, respectively. In the following discussion the
quantum numbers $\mathbf{k}$ and $\mathbf{q}$ will also denote implicitly
the left and right lead.

We model the AF on each side of the junction by a one-band Hubbard
Hamiltonian treated in the Hartree-Fock approximation for a static
spin-density wave (SDW) with wave-vector $\mathbf{Q}$. Without loss of
generality, we assume the SDW mean field to be polarized along the spin
quantization axis. Following Ref.~\cite{schrieffer}, we write 
\begin{equation}
\hat{H}_{L}=\sum_{\mathbf{k}\alpha }\epsilon _{k}c_{\mathbf{k}\alpha }^{\dag
}c_{\mathbf{k}\alpha }-\dfrac{US}{2}\sum_{\mathbf{k}\alpha \beta }c_{\mathbf{%
k}+\mathbf{Q}\alpha }^{\dag }\sigma _{\alpha \beta }^{3}c_{\mathbf{k}\beta
}\;,  \label{HF_Hubbard}
\end{equation}%
\noindent where $\epsilon _{k}$ is the band dispersion, $U$ the interaction
strength, $\sigma ^{3}$ the third Pauli matrix, while the order parameter $S$
is defined by $(1/N)\big\langle\sum_{\mathbf{k}\alpha \beta }c_{\mathbf{k}+%
\mathbf{Q}\alpha }^{\dag }\sigma _{\alpha \beta }^{3}c_{\mathbf{k}\beta }%
\big\rangle$ with $N$ the number of sites. This one-body Hamiltonian can be
diagonalized by the Bogoliubov transformation 
\begin{equation}
\begin{aligned} \gamma _{\mathbf{k}\alpha}^{c}
&=u_{\mathbf{k}}c_{\mathbf{k}\alpha}+v_{\mathbf{k}}\sum_{\beta }(\sigma
^{3})_{\alpha\beta }c_{\mathbf{k}+\mathbf{Q}\beta }\;, \label{Bogoliubov} \\
\gamma _{\mathbf{k}\alpha}^{v}
&=v_{\mathbf{k}}c_{\mathbf{k}\alpha}-u_{\mathbf{k}}\sum_{\beta }(\sigma
^{3})_{\alpha\beta }c_{\mathbf{k}+\mathbf{Q}\beta }\;. \end{aligned}
\end{equation}%
\noindent To avoid double counting, $\mathbf{k}$ is restricted to the
magnetic zone. The superscripts $c$ and $v$ refer to the conduction and the
valence bands split by the exchange Bragg scattering from the SDW. For
simplicity, we assume perfect nesting $\epsilon _{\mathbf{k}}=-\epsilon _{%
\mathbf{k}+\mathbf{Q}}$. In this case, the coefficients of the
transformation are given by $u_{\mathbf{k}}^{2}=\big[\tfrac{1}{2}\big(%
1+\epsilon _{\mathbf{k}}/E_{\mathbf{k}}\big)\big]$, $v_{\mathbf{k}}^{2}=\big[%
\tfrac{1}{2}\big(1-\epsilon _{\mathbf{k}}/E_{\mathbf{k}}\big)\big]$, $E_{%
\mathbf{k}}^{2}=(\epsilon _{\mathbf{k}}^{2}+\Delta ^{2})$, where $\Delta
=US/2$ is the SDW gap parameter \cite{half}. The diagonalized Hamiltonian is
given by $H=\sum_{\mathbf{k}\alpha }^{\ast }E_{\mathbf{k}}\big(\gamma _{%
\mathbf{k}\alpha }^{\dag c}\gamma _{\mathbf{k}\alpha }^{c}-\gamma _{\mathbf{k%
}\alpha }^{\dag v}\gamma _{\mathbf{k}\alpha }^{v}\big)$ where $\sum_{\mathbf{%
k}}^{\ast }$ means that the sum extends over the magnetic zone. The
single-particle energy spectrum is given by $\pm E_{\mathbf{k}}$ and the SDW
ground state for a half-filled band is defined by $\gamma _{\mathbf{k}\alpha
}^{\dag v}|\Omega \rangle =\gamma _{\mathbf{k}\alpha }^{c}|\Omega \rangle =0$%
, which may be found by filling the vacuum with valence-band quasiparticles
: 
\begin{align}
\left\vert \Omega \right\rangle & =\prod\limits_{\mathbf{k}\alpha }^{\ast
}(v_{\mathbf{k}}c_{\mathbf{k}\alpha }^{\dagger }-u_{\mathbf{k}}\sum_{\beta
}c_{\mathbf{k+Q}\beta }^{\dagger }\sigma _{\beta \alpha }^{3})\left\vert
0\right\rangle   \label{GS_standard} \\
& =\prod\limits_{\mathbf{k}\alpha }^{\ast }(v_{\mathbf{k}}-u_{\mathbf{k}%
}\sum_{\beta }c_{\mathbf{k+Q}\beta }^{\dagger }\sigma _{\beta \alpha }^{3}c_{%
\mathbf{k}\alpha })c_{\mathbf{k}\alpha }^{\dagger }\left\vert 0\right\rangle
.  \label{GS_BCS}
\end{align}%
The last form makes the analogies with the BCS ground state clear. For
example, there exists Andreev-like reflections at AF-N interfaces~\cite%
{Bobkova:2005,Nakamura:1999}. The ground state contains coherent
particle-hole pairs. To clarify this, we perform a particle-hole
transformation for states that are in the first magnetic Brillouin zone.
Recall that destroying an electron in a state, creates a hole in the
corresponding time-reversed state. For a spinor this can be achieved by $c_{%
\mathbf{k}\uparrow }\rightarrow h_{-\mathbf{k}\downarrow }^{\dagger }$ and $%
c_{\mathbf{k}\downarrow }\rightarrow -h_{-\mathbf{k}\uparrow }^{\dagger }.$
The ground state then takes the form 
\begin{equation}
\left\vert \Omega \right\rangle =\prod\limits_{\mathbf{k}}^{\ast }(v_{%
\mathbf{k}}-u_{\mathbf{k}}c_{\mathbf{k+Q}\uparrow }^{\dagger }h_{-\mathbf{k}%
\downarrow }^{\dagger })(v_{\mathbf{k}}-u_{\mathbf{k}}c_{\mathbf{k+Q}%
\downarrow }^{\dagger }h_{-\mathbf{k}\uparrow }^{\dagger })\left\vert
0\right\rangle _{h}
\end{equation}%
\noindent where $c_{\mathbf{k+Q}\alpha }\left\vert 0\right\rangle _{h}=0$
and $h_{\mathbf{k}\alpha }\left\vert 0\right\rangle _{h}=0$. The
particle-hole pair is in a triplet state with vanishing net spin projection
along the quantization axis, has no charge and has a wave vector equal to
the antiferromagnetic wave vector. In the case of a FM, that wave vector
would vanish.

In closing this section, note that for a N\'{e}el vector oriented in some
arbitrary direction with respect to the quantization axis, the variational
wave function Eq.(\ref{GS_BCS}) may be written%
\begin{equation}
\left\vert \Omega \right\rangle =\prod\limits_{\mathbf{k,}\alpha }^{\ast
}(v_{\mathbf{k}}-u_{\mathbf{k}}\sum_{\beta \delta \gamma }c_{\mathbf{k+Q}%
\delta }^{\dagger }U_{\delta \beta }\sigma _{\beta \gamma }^{3}U_{\alpha
\gamma }^{\dagger }c_{\mathbf{k}\gamma })\left\vert 0\right\rangle _{h}.
\end{equation}%
where 
\begin{equation}
U\left( \theta ,\phi \right) =\left( 
\begin{tabular}{ll}
$e^{-i\phi /2}\cos \left( \theta /2\right) $ & $-e^{-i\phi /2}\sin \left(
\theta /2\right) $ \\ 
$e^{i\phi /2}\sin \left( \theta /2\right) $ & $e^{i\phi /2}\cos \left(
\theta /2\right) $%
\end{tabular}%
\right) .  \label{Unitaire_def}
\end{equation}%
In a BCS superconductor, coherence of the pairs is reflected by the fact
that they all come with the same phase $e^{i\phi }$ in the BCS state $\prod_{%
\mathbf{k}}(u_{\mathbf{k}}-e^{i\phi }v_{\mathbf{k}}c_{\mathbf{k}\uparrow
}^{\dagger }c_{-\mathbf{k}\downarrow }^{\dagger })\left\vert 0\right\rangle
. $The $U\left( 1\right) $ rotation $e^{i\phi }$, identical for all pairs in
this case, has its analog in the $SU\left( 2\right) $ rotation $U_{\delta
\beta }\sigma _{\beta \gamma }^{3}U_{\alpha \gamma }^{\dagger }=\widehat{%
\mathbf{s}}\cdot \mathbf{\sigma }_{\delta \gamma },$with $\widehat{\mathbf{s}%
}\equiv \left( \sin \theta \cos \phi ,\sin \theta \sin \phi ,\cos \theta
\right) $ that, in the antiferromagnetic case, is applied to all
particle-hole pairs in the same way.

Although our results apply to any dimension, we often quote wave vectors for
two dimensions to simplify the discussion.

\subsection{Tunneling Hamiltonian}

We choose to define the spin quantization axis along the direction of the
instantaneous staggered magnetic moment $\mathbf{S}$. Since we are
interested in the case where the moments of the two AF are non-colinear (the
analog of a phase difference), the spin quantization axis will differ on
each side of the junction. It is thus necessary to include a unitary
transformation in spin space, defined by $\mathbf{U}(\theta ,\phi )$ in Eq.(%
\ref{Unitaire_def}) above, to account for the fact that a spin up on one
side of the junction is not the same as a spin up on the other side of the
junction. The angles $(\theta ,\phi )$ correspond to the orientation of $%
\mathbf{S}_{R}$ of the right AF expressed in the coordinate system of the
left side of the junction; in Cartesian coordinates $\mathbf{S}_{L}=|\mathbf{%
S}_{L}|(0,0,1)$ and $\mathbf{S}_{R}=|\mathbf{S}_{R}|(\sin \theta \cos \phi
,\sin \theta \sin \phi ,\cos \theta ).$ The annihilation operators transform
as follows%
\begin{equation}
d_{\mathbf{q\sigma }}=U_{\sigma \sigma ^{\prime }}\widetilde{d}_{\mathbf{q}%
\sigma ^{\prime }}  \label{Base_a_droite}
\end{equation}%
where $\widetilde{d}_{\mathbf{q}\sigma ^{\prime }}$ destroys a particle with
spin quantization axis along the direction of the N\'{e}el vector on the
right-hand side of the junction. The summation over repeated indices is
implied. The tunneling Hamiltonian then reads 
\begin{equation}
\hat{H}_{T}=(1/N)\sum_{\mathbf{k}\mathbf{q}\sigma \sigma ^{\prime }}\big(t_{%
\mathbf{k}\mathbf{q}}c_{\mathbf{k}\sigma }^{\dag }U_{\sigma \sigma ^{\prime
}}\widetilde{d}_{\mathbf{q}\sigma ^{\prime }}+h.c.\big).
\end{equation}%
The spin flip terms come purely from the choice of different quantization
axes on the left and on the right. There is no real spin flip in the
tunneling process.

By choosing the quantization axis along the N\'{e}el vector, we performed
the analog of a gauge choice. The choice of a different basis (quantization
axis) on either side is reflected completely in the $\mathbf{U}(\theta ,\phi
)$ appearing in the tunneling Hamiltonian. The angles in the unitary
transformation $\mathbf{U}(\theta ,\phi )$ are the non-Abelian analog of the
phase in the ordinary Josephson effect. In the latter case, electromagnetic
fields coupled to charge affect the phase difference $e^{-ie\xi /\hbar c}$
through a gauge transformation $A_{\mu }\rightarrow A_{\mu }+\partial _{\mu
}\xi $ on one side of the junction. Since we can write $\mathbf{U}(\theta
,\phi )=e^{-i\sigma _{z}\phi /2}e^{-i\sigma _{y}\theta /2}$ for example, we
see that in our case the gauge fields $\theta ,\phi $ are coupled to spins,
as is the magnetic field through the Zeeman effect. Such a field $\mathbf{B}$
yields a unitary time evolution $e^{ig\mu _{B}\mathbf{B\cdot \sigma }t/2}$
where where $g$ is the gyromagnetic ratio, $\mu _{B}$ is the Bohr magneton
and $t$ the time. Here it is the physical field and not the gauge field that
affects the generators of $SU\left( 2\right) $ rotations.

\section{Spin Josephson effect}

To obtain the usual Josephson effect in superconductivity, one computes the
time evolution of the number operator, which is conjugate to the phase. The
phase is the infinitesimal generator of the broken $U\left( 1\right) $
rotations. For a broken symmetry that is the staggered magnetization along $%
\widehat{\mathbf{z}}$, let us first consider the infinitesimal generator of $%
SU\left( 2\right) $ rotations along $x,$ $S^{x}\left( \mathbf{q}=0\right) .$%
Since the commutator $\left[ S^{x}\left( \mathbf{q}=0\right) ,S^{y}\left( 
\mathbf{q}=\left( \pi ,\pi \right) \right) \right] $ is proportional to the
order parameter, a constant, this means that the analog of the number
operator in this case is $S^{y}\left( \mathbf{q}=\left( \pi ,\pi \right)
\right) .$By considering $S^{y}\left( \mathbf{q}=0\right) $ for
infinitesimal rotations along $y,$one could have concluded that $S^{x}\left( 
\mathbf{q}=\left( \pi ,\pi \right) \right) $ was the analog of the number
operator \cite{Nogueira}. Furthermore, $S^{z}$ will change directions. This
means that one will find the analog of the Josephson effect, both $DC$ and $%
AC$, by writing down the equations of motion for the vector $\mathbf{S}%
\left( \mathbf{q}=\left( \pi ,\pi \right) \right) .$This is what we proceed
to do in this section.

\subsection{Steady state}

The staggered magnetic moment operator in the left lead is $\mathbf{\hat{S}}%
_{L}=(\hbar /2)\sum_{\mathbf{k}\alpha \beta }c_{\mathbf{k}+\mathbf{Q}\alpha
}^{\dag }\mathbf{\mathbf{\sigma }}_{\alpha \beta }c_{\mathbf{k}\beta }$
where $\mathbf{\sigma }$ is a vector of Pauli matrices. In the broken
symmetry state, the time evolution of $\mathbf{\hat{S}}_{L}$ due to $H_{L}$
is negligible. Since we also have $[\mathbf{\hat{S}}_{L},\hat{H}_{R}]=0$ we
find $d\mathbf{\hat{S}}_{L}/dt=(1/i\hbar )[\mathbf{\hat{S}}_{L},\hat{H}_{T}]$%
, and thus $d\mathbf{\hat{S}}_{L}/dt=-(i/2N)\sum_{\mathbf{k}\mathbf{q}%
}\sum_{\alpha \beta \delta }\big(\mathbf{\sigma }_{\alpha \beta }\mathbf{U}%
_{\beta \delta }\,t_{\mathbf{k}\mathbf{q}}c_{\mathbf{k}+\mathbf{Q}\alpha
}^{\dag }\widetilde{d}_{\mathbf{q}\delta }-h.c.\big)$ whose average $\dot{%
\mathbf{S}}_{L}(t)\equiv \langle d\mathbf{\hat{S}}_{L}/dt\rangle $ is given
by 
\begin{equation}
\dot{\mathbf{S}}_{L}(t)=\dfrac{1}{N}\sum_{\mathbf{k}\mathbf{q}}\sum_{\alpha
\beta \delta }\text{Im}\Big[\mathbf{\sigma }_{\alpha \beta }\mathbf{U}%
_{\beta \delta }\,t_{\mathbf{k}\mathbf{q}}\big\langle c_{\mathbf{k}+\mathbf{Q%
}\alpha }^{\dag }(t)\widetilde{d}_{\mathbf{q}\delta }(t)\big\rangle\Big]\;,
\label{Kubo1}
\end{equation}%
\noindent where $\langle ...\rangle $ is the thermal statistical average
with the full density matrix. Performing first order perturbation theory
using $\hat{H}_{T}$ as the perturbation, one obtains 
\begin{equation}
\big\langle c_{\mathbf{k}+\mathbf{Q}\alpha }^{\dag }\widetilde{d}_{\mathbf{q}%
\delta }\big\rangle=-\dfrac{i}{\hbar }\int_{-\infty }^{t}dt^{\prime }%
\big\langle\big[c_{\mathbf{k}+\mathbf{Q}\alpha }^{\dag }(t)\widetilde{d}_{%
\mathbf{q}\delta }(t),\hat{H}_{T}(t^{\prime })\big]\big\rangle_{0}\;,
\label{Kubo2}
\end{equation}%
\noindent where the average $\langle ...\rangle _{0}$ is computed with $%
H_{0}=\hat{H}_{L}+\hat{H}_{R}$ (the unperturbed part of $\hat{H}$). The
operators on the right are in the interaction representation.

Once the commutator is evaluated in Eq.~(\ref{Kubo2}), one of the
factorizations of the four-point correlation function involves products of
correlation functions on the left and on the right leads such as $\langle c_{%
\mathbf{k}+\mathbf{Q}\alpha }^{\dag }(t)c_{\mathbf{k}\alpha }(t^{\prime
})\rangle _{0}\langle \widetilde{d}_{\mathbf{q}+\mathbf{Q}\delta }(t)%
\widetilde{d}_{\mathbf{q}\delta }^{\dag }(t^{\prime })\rangle _{0}$. Such
correlation functions would vanish in a normal paramagnetic state. They are
non-zero because of the broken symmetry. They represent interference in the
tunneling process between momentum $\mathbf{k+Q}$ spin up particles and
momentum $-\mathbf{k}$ spin-down holes, in other words tunneling of charge
zero spin one $S^{z}=0$ coherent particle-hole pairs that have finite
momentum and are present in the ground-state Eq.(\ref{GS_BCS}). In the case
of the ordinary Josephson effect, one would find terms such as $\langle c_{%
\mathbf{k}\sigma }^{\dag }(t)c_{-\mathbf{k}-\sigma }^{\dag }(t^{\prime
})\rangle _{0}\langle \widetilde{d}_{\mathbf{q}\sigma }(t)\widetilde{d}_{-%
\mathbf{q}-\sigma }(t^{\prime })\rangle _{0}$ that represent tunneling of
coherent Cooper pairs.

In order to compute the averages $\langle...\rangle_0$ in the broken
symmetry states, we invert the Bogoliubov transformation Eq.~(\ref%
{Bogoliubov}). Assuming $t_{\mathbf{k} \mathbf{q}}=t_{\mathbf{k}\mathbf{q+Q}%
}=t_{\mathbf{k+Q}\mathbf{q}}=t_{\mathbf{k+Q}\mathbf{q+Q}}$, we find

\begin{equation}
\sum_{\mathbf{k}\mathbf{q}}t_{\mathbf{k}\mathbf{q}}c_{\mathbf{k}+\mathbf{Q}%
\alpha }^{\dag }\widetilde{d}_{\mathbf{q}\delta }=\sum_{\mathbf{k}\mathbf{q}%
}^{\ast }\sum_{i,j\in \{c,v\}}t_{\mathbf{k}\mathbf{q}}(\Gamma _{\mathbf{k}%
\mathbf{q}}^{\alpha \delta })_{ij}\gamma _{\mathbf{k}\alpha }^{i\dag }\gamma
_{\mathbf{q}\delta }^{j},
\end{equation}%
where we defined 
\begin{equation}
\begin{aligned} &(\Gamma^{\alpha\delta}_{\mathbf{k}\mathbf{q}})_{cc}\equiv
(u_{\mathbf{k}}u_{\mathbf{q}}+
\sigma^{3}_{\alpha\alpha}v_{\mathbf{k}}u_{\mathbf{q}}+\sigma^{3}_{\delta%
\delta}u_{\mathbf{k}}v_{\mathbf{q}} +
\sigma^{3}_{\alpha\alpha}\sigma^{3}_{\delta\delta}
v_{\mathbf{k}}v_{\mathbf{q}} )\\
&(\Gamma^{\alpha\delta}_{\mathbf{\mathbf{k}}\mathbf{q}})_{cv}\equiv
(v_{\mathbf{k}}u_{\mathbf{q}}-
\sigma^{3}_{\alpha\alpha}u_{\mathbf{k}}u_{\mathbf{q}}+\sigma^{3}_{\delta%
\delta}v_{\mathbf{k}}v_{\mathbf{q}} -
\sigma^{3}_{\alpha\alpha}\sigma^{3}_{\delta\delta}
u_{\mathbf{k}}v_{\mathbf{q}} )\\
&(\Gamma^{\alpha\delta}_{\mathbf{k}\mathbf{q}})_{vc}\equiv
(u_{\mathbf{k}}v_{\mathbf{q}}+
\sigma^{3}_{\alpha\alpha}v_{\mathbf{k}}v_{\mathbf{q}}-\sigma^{3}_{\delta%
\delta}u_{\mathbf{k}}u_{\mathbf{q}} -
\sigma^{3}_{\alpha\alpha}\sigma^{3}_{\delta\delta}
v_{\mathbf{k}}u_{\mathbf{q}} )\\
&(\Gamma^{\alpha\delta}_{\mathbf{k}\mathbf{q}})_{vv}\equiv
(v_{\mathbf{k}}v_{\mathbf{q}}-
\sigma^{3}_{\alpha\alpha}u_{\mathbf{k}}v_{\mathbf{q}}-\sigma^{3}_{\delta%
\delta}v_{\mathbf{k}}u_{\mathbf{q}} +
\sigma^{3}_{\alpha\alpha}\sigma^{3}_{\delta\delta}
u_{\mathbf{k}}u_{\mathbf{q}} )\;. \end{aligned}
\end{equation}

Similarly, let $\tilde{H}_{T}$ denote the part of $\hat{H}_{T}$ that does
not commute with $c_{\mathbf{k}+\mathbf{Q}\alpha }^{\dag }\widetilde{d}_{%
\mathbf{q}\delta }$. It can be written as $\tilde{H}_{T}=(1/N)\sum_{\mathbf{k%
}\mathbf{q}\sigma \delta ^{\prime }}^{\ast }\sum_{ij}\mathbf{U}_{\sigma
\delta ^{\prime }}^{\ast }t_{\mathbf{k}\mathbf{q}}^{\ast }(\Gamma _{\mathbf{k%
}\mathbf{q}}^{\sigma \delta ^{\prime }})_{ij}\gamma _{\mathbf{q}\delta
^{\prime }}^{j\dag }\gamma _{\mathbf{k}\sigma }^{i}$. Substituting these
expressions into Eq. (\ref{Kubo2}) and taking into account the fact that to
this order in $t_{\mathbf{k}\mathbf{q}}$ the unitary transformation $\mathbf{%
U}\left( t^{\prime }\right) $ in the tunneling matrix element can be
evaluated at $t^{\prime }=t$ since $\hat{H}_{L}$ and $\hat{H}_{R}$ do not
change the quantization axis$,$ one finds 
\begin{widetext}
\begin{equation}\label{Js_G_omega}
\begin{aligned}
                \dot{\mathbf{S}}_{L}(t)
	&=\dfrac{1}{N^2}\sum^{*}_{\mathbf{k}\mathbf{q}}\sum_{\alpha\beta\delta}\sum_{\sigma\delta'}\text{Im}\Bigg[-\dfrac{i}{\hbar}
	|t_{\mathbf{k}\mathbf{q}}|^{2}\vec{\sigma}_{\alpha\beta}
	\mathbf{U}_{\beta\delta}\mathbf{U}^{*}_{\sigma\delta'}\int dt' e^{-0^{+}(t-t')}\times\\
	&\qquad\qquad\qquad\qquad\qquad\qquad\sum_{ij}
	(\Gamma^{\alpha\delta}_{\mathbf{k}\mathbf{q}})_{ij}(\Gamma^{\sigma\delta'}_{\mathbf{k}\mathbf{q}})_{ij}
	\Big(\mathcal{G}^{i<}_{\mathbf{k}\sigma\alpha}(t'-t)
	\mathcal{G}^{j>}_{\mathbf{q}\delta\delta'}(t-t') -
	\mathcal{G}^{i>}_{\mathbf{k}\sigma\alpha}(t'-t)\mathcal{G}^{j<}_{\mathbf{q}\delta\delta'}(t-t')\Big)\Bigg]
	\end{aligned}
	\end{equation}
\end{widetext}where $\mathcal{G}_{\mathbf{k}(\mathbf{q})}^{i<(>)}$ are the
Keldysh Green functions in the left (right) lead. Their definitions are $%
\mathcal{G}_{\mathbf{k}(\mathbf{q}),\alpha \beta }^{i<}(t,t^{\prime
})=i\langle \gamma _{\mathbf{k}(\mathbf{q})\beta }^{i\dag }(t^{\prime
})\gamma _{\mathbf{k}(\mathbf{q})\alpha }^{i}(t)\rangle $ and $\mathcal{G}_{%
\mathbf{k}(\mathbf{q}),\alpha \beta }^{i>}(t,t^{\prime })=-i\langle \gamma _{%
\mathbf{k}(\mathbf{q})\alpha }^{i}(t)\gamma _{\mathbf{k}(\mathbf{q})\beta
}^{i\dag }(t^{\prime })\rangle $, respectively. Explicitly, 
\begin{align*}
\mathcal{G}_{\mathbf{k}\sigma \sigma ^{\prime }}^{i>}(t^{\prime }-t)&
=-i(1-f(E_{k}^{i}))\exp [-iE_{k}^{i}(t^{\prime }-t)/\hbar ]\delta _{\sigma
\sigma ^{\prime }}\;, \\
\mathcal{G}_{\mathbf{k}\sigma \sigma ^{\prime }}^{i<}(t^{\prime }-t)&
=if(E_{k}^{i})\exp [-iE_{k}^{i}(t^{\prime }-t)/\hbar ]\delta _{\sigma \sigma
^{\prime }}\;,
\end{align*}%
where $f$ is the Fermi function. Eq.~(\ref{Js_G_omega}) for the staggered
magnetic moment current through a tunnel junction is general. A bias could
be included. We assume that there is no bias so there is no incoherent
single-particle tunneling across the antiferromagnetic gap. Integrating over 
$t-t^{\prime }$ and performing the spin sum in (\ref{Js_G_omega}), one finds 
\begin{equation}
\dot{\mathbf{S}}_{L}=I_{c}\,\hat{\mathbf{s}}_{L}\times \hat{\mathbf{s}}%
_{R}\;,  \label{mainresult1}
\end{equation}%
\noindent where $\hat{\mathbf{s}}_{L(R)}=\mathbf{S}_{L(R)}/|\mathbf{S}%
_{L(R)}|.$To obtain the correct sign, one must take into account in the
Fourier transforms that the last site to the left and the first site to the
right do not belong to the same sublattice. In the above equation, $I_{c}$
is defined by 
\begin{equation*}
I_{c}=\dfrac{8\Delta _{L}\Delta _{R}}{N^{2}}P\sum_{kq}^{\ast }|t_{\mathbf{kq}%
}|^{2}\dfrac{f(E_{\mathbf{k}})-f(-E_{\mathbf{k}})}{E_{\mathbf{k}}(E_{\mathbf{%
k}}^{2}-E_{\mathbf{q}}^{2})}
\end{equation*}%
with $P$ indicating principal part. A similar expression is found for the
equilibrium spin current in the case of ferromagnetic tunnel junctions. Note
that the sine function present in the standard Josephson case is replaced
here by a cross product, which is a direct consequence of the vectorial
nature of the order parameter.

For a symmetrical junction ($\Delta _{L}=\Delta _{R}$), the same assumptions
and procedure as Ref.~\cite{ambegaokar} lead to the following analytical
result 
\begin{equation}
I_{c}=\dfrac{h}{e^{2}}R^{-1}\Delta (T)\tanh (\tfrac{1}{2}\beta \Delta (T))\;,
\end{equation}%
\noindent where $R=\hbar /\left( 4\pi e^{2}D^{2}\left\vert t\right\vert
^{2}\right) $ is the (zero-temperature) normal-state resistance of the
junction with $D$ the density of state, which is assumed to be a constant.
This expression for the temperature dependence of the critical current has
the same form as that obtained by Ambegaokar and Baratoff \cite{ambegaokar}
for a BCS superconductor, which is not surprising given the formal analogies 
\cite{Over}.

By symmetry, the time derivative of the staggered magnetic moment on the
right lead can be obtained by interchanging the $L$ and $R$ indices in Eq.~(%
\ref{mainresult1}). As a consequence, the staggered magnetic moments of the
two AF precess about their (constant) sum $\mathbf{S}_{L}+\mathbf{S}_{R}$ at
a frequency $\omega _{0}=I_{c}|\mathbf{S}_{L}+\mathbf{S}_{R}|/|\mathbf{S}%
_{L}||\mathbf{S}_{R}|$.

\subsection{AC spin Josephson effect}

In the ordinary Josephson effect, the electromagnetic gauge potentials enter
directly in the argument of the sine function. The present case is
different. Each magnetic moment associated with a spin couples to the
magnetic field through the Zeeman term ($H_{Z}=-g\mu _{B}\mathbf{B\cdot S}$)
where $g$ is the gyromagnetic ratio and $\mu _{B}$ the Bohr magneton (we
neglect terms coming from orbital motion \cite{Bulaevskii:2008}).

Considering magnetic fields $\mathbf{B}_{L}$ and $\mathbf{B}_{R}$ applied
respectively to the left- and right-hand sides of the junction, the
Heisenberg equations of motion lead to the following equations of motion for
the order parameters 
: 
\begin{align}
\dot{\mathbf{S}}_{L}& =-g\mu _{B}\mathbf{B}_{L}\times \mathbf{S}_{L}+I_{c}\,%
\hat{\mathbf{s}}_{L}\times \hat{\mathbf{s}}_{R},  \notag \\
\dot{\mathbf{S}}_{R}& =-g\mu _{B}\mathbf{B}_{R}\times \mathbf{S}_{R}+I_{c}\,%
\hat{\mathbf{s}}_{R}\times \hat{\mathbf{s}}_{L}\;.  \label{dS/dt_AC}
\end{align}%
The first term on the right side of the equality is merely the contribution
of $H_{Z}$ to the Heisenberg equation of motion. The second term is the
tunneling contribution and has exactly the same form as the one we have
already computed in the zero-field case, namely Eq.(\ref{mainresult1}).

To prove the correctness of the $I_{c}\hat{\mathbf{s}}_{L}\times \hat{%
\mathbf{s}}_{R}$ terms, we must return to the contribution from the first
order in perturbation theory term Eq.(\ref{Kubo2}). The expectation values
are now evaluated with the unperturbed Hamiltonian $\hat{H}_{L}-g\mu _{B}%
\mathbf{B}_{L}\mathbf{\cdot S}+\hat{H}_{R}-g\mu _{B}\mathbf{B}_{R}\mathbf{%
\cdot S.}$ Contrary to the case where no magnetic field was applied, the
quantization axes now precess and one must worry about the time dependence
of the rotation matrix $U_{\delta ^{\prime }\sigma ^{\prime }}^{\dag
}(t^{\prime })$ in terms of the form $\widetilde{d}_{\mathbf{q}\delta
^{\prime }}^{\dag }(t^{\prime })U_{\delta ^{\prime }\sigma ^{\prime }}^{\dag
}(t^{\prime })c_{\mathbf{k}\sigma ^{\prime }}(t^{\prime })$. That rotation
matrix is no longer evaluated at time $t.$ The creation-annihilation
operators are in the interaction representation and the original Hamiltonian
commutes with the Zeeman Hamiltonian. Hence, it is possible to factorize the
time evolution due to the magnetic fields to obtain $\widetilde{d}_{\mathbf{q%
}\delta ^{\prime }}^{\dag }(t^{\prime })=\widetilde{d}_{\mathbf{q}\delta
}^{0\dag }(t^{\prime })\Lambda _{\delta \delta ^{\prime }}^{R\dag }\left(
t^{\prime }\right) $ and $c_{\mathbf{k}\sigma ^{\prime }}(t^{\prime
})=\Lambda _{\sigma ^{\prime }\sigma }^{L}\left( t^{\prime }\right) c_{%
\mathbf{k}\sigma }^{0}(t^{\prime })$ where 
\begin{equation}
\Lambda _{\sigma ^{\prime }\sigma }^{R(L)}\left( t^{\prime }\right) =\exp
[ig\mu _{B}\mathbf{B}_{R(L)}\cdot \mathbf{\sigma }t^{\prime }/2]_{\sigma
^{\prime }\sigma }
\end{equation}%
and the superscript $0$ on creation-annihilation operators indicates time
evolution with $\ \hat{H}_{0}=\hat{H}_{L}+\hat{H}_{R}$. Since $\Lambda
_{\sigma ^{\prime }\sigma }^{L}\left( t^{\prime }\right) =\Lambda _{\sigma
^{\prime }\sigma }^{L}\left( t^{\prime }-t\right) \Lambda _{\sigma \overline{%
\sigma }}^{L}\left( t\right) $ we can also write 
\begin{equation}
c_{\mathbf{k}\sigma ^{\prime }}(t^{\prime })=\Lambda _{\sigma ^{\prime
}\sigma }^{L}\left( t^{\prime }-t\right) \Lambda _{\sigma \overline{\sigma }%
}^{L}\left( t\right) c_{\mathbf{k}\overline{\sigma }}^{0}(t^{\prime })
\end{equation}%
and define creation-annihilation operators whose spin basis is rotated by
the magnetic field, 
\begin{equation*}
c_{\mathbf{k}\sigma }^{t}(t^{\prime })\equiv \Lambda _{\sigma \overline{%
\sigma }}^{L}\left( t\right) c_{\mathbf{k}\overline{\sigma }}^{0}(t^{\prime
})
\end{equation*}%
so that 
\begin{equation}
c_{\mathbf{k}\sigma ^{\prime }}(t^{\prime })=\Lambda _{\sigma ^{\prime
}\sigma }^{L}\left( t^{\prime }-t\right) c_{\mathbf{k}\sigma }^{t}(t^{\prime
}).  \label{Transfo_ct}
\end{equation}%
The superscript $t$ in $c_{\mathbf{k}\sigma }^{t}(t^{\prime })$ reminds us
that the spin basis is that defined by the N\'{e}el vector at time $t.$ That
basis rotates because of the magnetic field. Somewhat redundantly, we also
use $\sigma $ and $\sigma ^{\prime }$ to denote the spin basis on the left
of the junction at time $t$ and $t^{\prime }$ respectively. Physically, the
last result tells us that the time evolution of the basis due to the
magnetic field can be taken into account separately.

We now need to work out the corresponding transformation laws for the
unitary transformation $U_{\delta ^{\prime }\sigma ^{\prime }}^{\dag
}(t^{\prime }).$This operator changes the basis from left $\left( \sigma
^{\prime }\right) $ to right $\left( \delta ^{\prime }\right) $ at time $%
t^{\prime }.$ The only time dependence in $U_{\delta ^{\prime }\sigma
^{\prime }}^{\dag }(t^{\prime })$ comes from that of the basis on the left
and on the right due to the magnetic fields, it is not an additional time
dependence. Hence, the transformation law, which we prove in more details
below, is 
\begin{equation}
U_{\delta ^{\prime }\sigma ^{\prime }}^{\dag }(t^{\prime })=\Lambda _{\delta
^{\prime }\delta }^{R}\left( t^{\prime }-t\right) U_{\delta \sigma }^{\dag
}(t)\Lambda _{\sigma \sigma ^{\prime }}^{L\dagger }\left( t^{\prime
}-t\right) .  \label{Transfo_des_unitaires}
\end{equation}%
where we let $\delta $ and $\delta ^{\prime }$ denote the spin basis on the
right at times $t$ and $t^{\prime }$respectively. Given the transformation
law for the operators on the left Eq.(\ref{Transfo_ct}) and the analogous
one for the operators on the right, we will thus find 
\begin{equation}
\widetilde{d}_{\mathbf{q}\delta ^{\prime }}^{\dag }(t^{\prime })U_{\delta
^{\prime }\sigma ^{\prime }}^{\dag }(t^{\prime })c_{\mathbf{k}\sigma
^{\prime }}(t^{\prime })=\widetilde{d}_{\mathbf{q}\delta }^{t\dag
}(t^{\prime })U_{\delta \sigma }^{\dag }(t)c_{\mathbf{k}\sigma
}^{t}(t^{\prime })  \label{Invariance}
\end{equation}%
i.e. all spin indices are at time $t$ on the right-hand side of the
equality. Therefore, it is possible to factor the unitary transformation $%
U_{\delta \sigma }^{\dag }(t)$ out of the integral over $t^{\prime }$ in the
perturbation formula Eq.(\ref{Kubo2}). The rest of the calculation then just
follows the same path as in the zero-field case, leading to the above
equations of motion Eq.(\ref{dS/dt_AC}). 

We now complete the proof on the transformation properties of the unitary
matrices in Eq.(\ref{Transfo_des_unitaires}). First, express the left-hand
side creation operators alternatively in the right and in the left-hand side
basis. More specifically, left multiplying by $U^{\dag }$ the change of
basis Eq.(\ref{Base_a_droite}), the annihilation operators of left-hand side
can be expressed in the spin basis of the right-hand side%
\begin{equation}
U_{\delta ^{\prime }\sigma ^{\prime }}^{\dag }\left( t^{\prime }\right) c_{%
\mathbf{k}\sigma ^{\prime }}(t^{\prime })=\widetilde{c}_{\mathbf{k}\delta
^{\prime }}(t^{\prime }).
\end{equation}%
Now we change to the right-hand side basis at time $t$ with Eq.(\ref%
{Transfo_ct})%
\begin{equation}
U_{\delta ^{\prime }\sigma ^{\prime }}^{\dag }\left( t^{\prime }\right) c_{%
\mathbf{k}\sigma ^{\prime }}(t^{\prime })=\Lambda _{\delta ^{\prime }\delta
}^{R}\left( t^{\prime }-t\right) \widetilde{c}_{\mathbf{k}\delta
}^{t}(t^{\prime })
\end{equation}%
and we return to the creation operators in the spin basis appropriate for
the left-hand side at time $t$.%
\begin{equation}
U_{\delta ^{\prime }\sigma ^{\prime }}^{\dag }\left( t^{\prime }\right) c_{%
\mathbf{k}\sigma ^{\prime }}(t^{\prime })=\Lambda _{\delta ^{\prime }\delta
}^{R}\left( t^{\prime }-t\right) U_{\delta \sigma }^{\dag }\left( t\right) 
\widetilde{c}_{\mathbf{k}\sigma }^{t}(t^{\prime }).
\end{equation}%
We arrive at our final result by inserting the identity and changing the
spin basis from time $t$ to time $t^{\prime },$ using Eq.(\ref{Transfo_ct})
again 
\begin{widetext}
\begin{eqnarray}
U_{\delta ^{\prime }\sigma ^{\prime }}^{\dag }\left( t^{\prime }\right) c_{%
\mathbf{k}\sigma ^{\prime }}(t^{\prime }) &=&\Lambda _{\delta ^{\prime
}\delta }^{R}\left( t^{\prime }-t\right) U_{\delta \sigma }^{\dag }\left(
t\right) \left( \Lambda _{\sigma \sigma ^{\prime }}^{\dag L}\left( t^{\prime
}-t\right) \Lambda _{\sigma ^{\prime }\alpha }^{L}\left( t^{\prime
}-t\right) \right) \widetilde{c}_{\mathbf{k}\alpha }^{t}(t^{\prime }) \\
&=&\left( \Lambda _{\delta ^{\prime }\delta }^{R}\left( t^{\prime }-t\right)
U_{\delta \sigma }^{\dag }\left( t\right) \Lambda _{\sigma \sigma ^{\prime
}}^{\dag L}\left( t^{\prime }-t\right) \right) c_{\mathbf{k}\sigma ^{\prime
}}(t^{\prime }).
\end{eqnarray}%
\end{widetext}%
Comparing the left and right-hand sides, this completes the proof of the
transformation property for the change of basis from left to right, Eq.(\ref%
{Invariance}) at different times.

We close this section by solving and interpreting the solution of the
equations of motion Eq.(\ref{dS/dt_AC}). If $\mathbf{B}_{L}=\mathbf{B}_{R}=%
\mathbf{B}$, one finds that in the rotating frame $\mathbf{\hat{u}}=\{%
\mathbf{\hat{x}}^{\prime },\mathbf{\hat{y}}^{\prime },\mathbf{\hat{z}}%
^{\prime }\}$ defined by $d\mathbf{\hat{u}}/dt=-g\mu _{B}\mathbf{B}\times 
\mathbf{\hat{u}}$, $\mathbf{S}_{L}$ and $\mathbf{S}_{R}$ still precess about
their (constant) sum $\mathbf{\Sigma }\equiv \mathbf{S}_{L}+\mathbf{S}_{R}$
at a frequency $\omega _{0}$. Returning to the static frame, Eq.(\ref%
{dS/dt_AC}) gives $d\mathbf{\Sigma }/dt=-g\mu _{B}\mathbf{B}\times \mathbf{%
\Sigma }$ so that $\mathbf{S}_{L}$ and $\mathbf{S}_{R}$ undergo a motion of
double precession.

In the ordinary Josephson effect, a constant electric potential difference $%
V $ leads to a phase difference $\varphi $ that depends linearly on time ($%
\dot{\varphi}=2eV/\hbar $). In the present case, an analog is found by
computing the time dependence of the relative orientation $\vartheta $
between $\mathbf{S}_{L}$ and $\mathbf{S}_{R}$. Using Eq. (\ref{dS/dt_AC}) to
compute $d(\cos \vartheta )/dt=d(\hat{\mathbf{s}}_{L}\cdot \hat{\mathbf{s}}%
_{R})/dt$, we obtain 
\begin{equation}
\dot{\vartheta}(t)=-g\mu _{B}\delta \mathbf{B}\cdot \hat{\mathbf{e}}(t),
\label{AC_spin_Josephson}
\end{equation}%
where $\delta \mathbf{B}\equiv \mathbf{B}_{L}-\mathbf{B}_{R}$ and $\hat{%
\mathbf{e}}(t)\equiv \hat{\mathbf{s}}_{R}(t)\times \hat{\mathbf{s}}_{L}(t)$.
As in the ordinary Josephson effect, this expression does not contain
explicitly the tunneling matrix element. Contrary to the ordinary Josephson
effect, this equation is non-linear. Solving it numerically along with Eq. (%
\ref{dS/dt_AC}) in the case where the magnetic field vanishes on one side of
the junction, one finds that the angle between $\mathbf{S}_{L}$ and $\mathbf{%
S}_{R}$ behaves as a sine-like function of time. The presence of an
additional constant magnetic field $\mathbf{B}$ throughout the system adds a
beat to this sine-like behavior. The gyromagnetic ratio $g$ is material
dependent. No such non-universal constant appears in the ordinary Josephson
effect. The above discussion of the AC effect can be transposed for FM by
replacing the staggered magnetic moment by the uniform one.


\section{Experimental detection}

In principle, to detect the effects described above, one could use
antiferromagnetic resonance\cite{Kittel}. This phenomenon is based on an
anisotropy field resulting from the spin-orbit interaction. However the
effect is absent for a spin $1/2.$ Assuming however that the same equations
would hold for $S=1,$ it becomes possible to detect the effect by
antiferromagnetic resonance \cite{Kittel}. In the limit where $\mathbf{S}%
_{L} $ and $\mathbf{S}_{R}$ are nearly colinear, the effect should therefore
lead to a uniform shift of order $\omega _{0}$ in the antiferromagnetic
resonance frequencies of each AF. 
With $|\tfrac{2\Delta }{S}|=U=2\text{eV}$ and a normal state conductance of
the order of the conductance quantum, $R^{-1}=2e^{2}/h$, one finds a value
of $\omega _{0}$ of order $10^{14}$ Hz, in other words in the visible. This
frequency would be higher than the single particle gap and so would lead to
much damping. Resistances that are many orders of magnitude larger are thus
needed to bring its value down.

To detect the spin Josephson effect even in the spin $1/2$ case, we suggest
to add an additional tunnel junction to a ferromagnet. In other words,
consider an AF/I/AF/I/FM junction. Indeed, let us begin just with the
AF/I/FM junction. The ferromagnetic material can be described by the Stoner
Hamiltonian%
\begin{equation}
H=\left( 
\begin{array}{cc}
d_{\mathbf{q\uparrow }}^{\dagger } & d_{\mathbf{q\downarrow }}^{\dagger }%
\end{array}%
\right) \left( 
\begin{array}{cc}
\zeta _{\mathbf{q}}-h & 0 \\ 
0 & \zeta _{\mathbf{q}}+h%
\end{array}%
\right) \left( 
\begin{array}{c}
d_{\mathbf{q\uparrow }} \\ 
d_{\mathbf{q\downarrow }}%
\end{array}%
\right) 
\end{equation}%
where $\zeta _{\mathbf{q}}=\varepsilon _{\mathbf{q}}-\mu $ and $h$ is a
molecular field. Using the same tunneling Hamiltonian as before and
following the same procedure, we find that 
\begin{eqnarray}
\mathbf{\dot{S}}_{L} &=&\mathcal{I}_{c}\mathbf{M}_{R}\times \mathbf{S}_{L} \\
\mathbf{\dot{M}}_{R} &=&\mathcal{I}_{c}\mathbf{S}_{L}\times \mathbf{M}_{R}
\end{eqnarray}%
where $\mathbf{M}_{R}$ is the average magnetization of the ferromagnet and 
\begin{eqnarray}
\mathcal{I}_{c} &=&\frac{1}{N^{2}}\mathcal{P}\sum_{\mathbf{k}}^{\ast }\sum_{%
\mathbf{q}}\left\vert t_{\mathbf{kq}}\right\vert ^{2}\frac{\Delta _{L}}{E_{%
\mathbf{k}}}\left\{ \frac{f\left( \zeta _{\mathbf{q}}-h\right) -f\left( E_{%
\mathbf{k}}\right) }{\zeta _{\mathbf{q}}-h-E_{\mathbf{k}}}\right.  \\
&&\left. -\frac{f\left( \zeta _{\mathbf{q}}-h\right) -f\left( -E_{\mathbf{k}%
}\right) }{\zeta _{\mathbf{q}}-h+E_{\mathbf{k}}}-\left( h\rightarrow
-h\right) \right\} .
\end{eqnarray}%
Since the uniform magnetization of the ferromagnet precesses due to the N%
\'{e}el order parameter, it should be possible to detect the precession of
the FM moment by standard magnetic resonance experiments. The addition of a
second AF junction, as suggested above, should affect the ferromagnetic
resonance.

\section{Generalized Josephson effects between systems where
antiferromagnetism and $d$-wave superconductivity coexist}

In this section, we apply the same approach to study tunneling currents
across a junction between two materials that are d-wave superconducting
antiferromagnets, i.e. with homogeneous coexistence of antiferromagnetism
and d-wave superconductivity. Since in the end we expect symmetry
considerations instead of details of the Hamiltonian to be the dominant
effect, we start with the simplest mean-field model that yields coexistence
of the two order parameters. Then we show that for coexisting orders, the
ordinary Josephson effect can be modulated by the antiferromagnetism and
that the spin Josephson effect can be modulated by the superconductivity.
These effects occur already to leading order in the square of the tunneling
matrix element.

\subsection{Model}

To take the simplest possible case that has the appropriate symmetries, we
model the AF/SC coexistence on each side of the junction by the following
phenomenological mean-field Hamiltonian \cite{bumsoo} : 
\begin{align}
\hat{H}=& \sum_{\mathbf{k}\sigma }(\xi _{\mathbf{k}}-\mu )c_{\mathbf{k}%
\sigma }^{\dag }c_{\mathbf{k}\sigma }+U\sum_{i}(c_{i,\uparrow }^{\dag
}c_{i,\uparrow }\langle c_{i,\downarrow }^{\dag }c_{i,\downarrow }\rangle
+h.c.)  \notag \\
& -V\sum_{i}(\Delta _{d,i}^{\dag }\langle \Delta _{d,i}\rangle
+h.c.)-W\sum_{i}(\Delta _{t,i}^{\dag }\langle \Delta _{t,i}\rangle +h.c.)\;.
\end{align}%
%
%
%
%
%
%
%
%
%
%
%
%
%
%
%
%
%
%
%
%
%
%
The possibility of a spin-triplet pair amplitude $\langle \Delta
_{t,i}\rangle $ must also be considered in general for self-consistency \cite%
{w_egal_0}. The singlet $\Delta _{d,i}$ and triplet $\Delta _{t,i}$ pair
destruction operators are defined as 
\begin{align}
\Delta _{d,i}& =\dfrac{1}{2}\sum_{\delta }g(\delta )(c_{i+\delta ,\uparrow
}c_{i,\downarrow }-c_{i+\delta ,\downarrow }c_{i,\uparrow }), \\
\Delta _{t,i}& =\dfrac{1}{2}\sum_{\delta }g(\delta )(c_{i+\delta ,\uparrow
}c_{i,\downarrow }+c_{i+\delta ,\downarrow }c_{i,\uparrow })
\end{align}%
and the structure factor $g(\delta )$ is chosen to have a d-wave-like form
such as 
\begin{equation}
g(\delta )=%
\begin{cases}
1/2\;\;\;\text{if}\;\;\;\delta =(\pm 1,0)\;, \\ 
-1/2\;\;\;\text{if}\;\;\;\delta =(0,\pm 1)\;, \\ 
0\;\;\;\text{if otherwise}\;.%
\end{cases}%
\end{equation}

The strength of AF, SC, and spin-triplet pair interactions is governed by $U$%
, $V$, and $W$, respectively, which are all positive in our study. $\xi _{%
\mathbf{k}}$ is a tight binding energy dispersion given as $\xi _{\mathbf{k}%
}=-2t(\cos k_{x}+\cos k_{y})-4t^{\prime }\cos k_{x}\cos k_{y}$, where $t$
and $t^{\prime }$ are hopping constants for nearest neighbors and next
nearest neighbors, and $\mu $ is the chemical potential, which controls the
particle density $n$.

For completeness, we recall the results of Ref. 
\onlinecite{bumsoo}
that we need. In the mean-field approximation, three different order
parameters $m$, $s$, and $t$ corresponding to the three different
interactions may be defined in the following way: 
\begin{align}
& \langle c_{i\sigma }^{\dag }c_{i\sigma }\rangle =\langle n_{i\sigma
}\rangle =\dfrac{n}{2}+\sigma m\cos (\mathbf{Q}\cdot \mathbf{r}_{i})\;, \\%
[12pt]
& \dfrac{1}{2}\sum_{\delta }g(\delta )\langle c_{i+\delta ,\uparrow
}c_{i,\downarrow }-c_{i+\delta ,\downarrow }c_{i,\uparrow }\rangle =s\;, \\%
[12pt]
\dfrac{1}{2}\sum_{\delta }& g(\delta )\langle c_{i+\delta ,\uparrow
}c_{i,\downarrow }+c_{i+\delta ,\downarrow }c_{i,\uparrow }\rangle =t\cos (%
\mathbf{Q}\cdot \mathbf{r}_{i})\;,  \label{t_order_parameter}
\end{align}%
where $\mathbf{Q}$ is the (commensurate) antiferromagnetic wave vector equal
to $(\pi ,\pi )$ in two dimensions. Now the mean-field Hamiltonian $H_{\text{%
MF}}$ is quadratic in the original electron operator and in terms of a new
four component field operator $\psi _{\mathbf{k}}$ it becomes bilinear 
\begin{equation}
\hat{H}_{CM}=\sum_{\mathbf{k}}^{\ast }\psi _{\mathbf{k}}^{\dag }M_{\mathbf{k}%
}\psi _{\mathbf{k}}+E_{0}\;,  \label{H_CM}
\end{equation}%
where 
\begin{equation}
\psi _{\mathbf{k}}^{\dag }\equiv \big(c_{\mathbf{k}\uparrow }^{\dag },c_{-%
\mathbf{k}\downarrow },c_{\mathbf{k}+\mathbf{Q}\uparrow }^{\dag },c_{-%
\mathbf{k}-\mathbf{Q}\downarrow }\big).
\end{equation}%
The matrix $M_{\mathbf{k}}$ is given as 
\begin{equation}
M_{\mathbf{k}}=%
\begin{pmatrix}
\epsilon _{\mathbf{k}} &  & Vs\phi (\mathbf{k}) &  & -Um &  & Wt\phi (%
\mathbf{k}) \\ 
Vs\phi (\mathbf{k}) &  & -\epsilon _{\mathbf{k}} &  & -Wt\phi (\mathbf{k}) & 
& -Um \\ 
-Um &  & -Wt\phi (\mathbf{k}) &  & \epsilon _{\mathbf{k}+\mathbf{Q}} &  & 
-Vs\phi (\mathbf{k}) \\ 
Wt\phi (\mathbf{k}) &  & -Um &  & -Vs\phi (\mathbf{k}) &  & -\epsilon _{%
\mathbf{k}+\mathbf{Q}}%
\end{pmatrix}%
\;,  \label{matrice_M}
\end{equation}%
where 
\begin{equation}
\epsilon _{\mathbf{k}}=\xi _{\mathbf{k}}-\mu
\end{equation}%
and $\phi (\mathbf{k})=\cos k_{x}-\cos k_{y}$ is the Fourier transform of $%
g(\delta )$. The constant energy shift $E_{0}$ depending on $m$, $s$, $t$ is
given as 
\begin{equation}
E_{0}=N(Um^{2}+Vs^{2}+Wt^{2}-\mu )\;,
\end{equation}%
where $N$ is the total number of lattice sites. The energy eigenvalues of $%
M_{\mathbf{k}}$ yield four energy dispersions $\pm E_{\pm }(\mathbf{k})$
(with $E_{\pm }(\mathbf{k})>0$) 
\onlinecite{bumsoo}%
\onlinecite{Murakami:1998}%
, 
\begin{align}
E_{\pm }(\mathbf{k})=& \{(\epsilon _{\mathbf{k}}^{2}+\epsilon _{\mathbf{k}+%
\mathbf{Q}}^{2})/2+(Um)^{2}+[Vs\phi (\mathbf{k})]^{2}  \notag \\
& +[Wt\phi (\mathbf{k})]^{2}\pm g(\mathbf{k})\}^{1/2}\;,
\end{align}%
where $g(\mathbf{k})$ is given as 
\begin{align}
g(\mathbf{k})=& \{(\epsilon _{\mathbf{k}}^{2}-\epsilon _{\mathbf{k}+\mathbf{Q%
}}^{2})^{2}/4+(\epsilon _{\mathbf{k}}-\epsilon _{\mathbf{k}+\mathbf{Q}%
})^{2}[Wt\phi (\mathbf{k})]^{2}  \notag \\
& +\big((\epsilon _{\mathbf{k}}+\epsilon _{\mathbf{k}+\mathbf{Q}%
})(Um)+2[Vs\phi (\mathbf{k})][Wt\phi (\mathbf{k})]\big)^{2}\}^{1/2}\;.
\end{align}%
When $s=t=0$ or $m=t=0$ the energy eigenvalues reduce to those obtained from
the SDW or BCS factorizations of the corresponding interactions.


The transformation matrix $A$ relating the original $c$ operator to the
eigenoperator $\gamma $ is defined by 
\begin{equation}
\psi _{i\mathbf{k}}=\sum_{j}A_{ij}(\mathbf{k})\gamma _{j\mathbf{k}}\;.
\end{equation}%
All the elements of the matrix $A$ are given in the appendix of Ref.\ 
\onlinecite{bumsoo}%
. The columns of the matrix $A$ are the eigenvectors of the mean-field
Hamiltonian with the corresponding eigenenergies ordered as follows, $E_{%
\mathbf{k}}^{1}=+E_{+}(\mathbf{k})$, $E_{\mathbf{k}}^{2}=-E_{+}(\mathbf{k})$%
, $E_{\mathbf{k}}^{3}=-E_{-}(\mathbf{k})$, $E_{\mathbf{k}}^{4}=+E_{-}(%
\mathbf{k})$. The vacuum state of $\psi _{\mathbf{k}}$ is obtained by
filling the vacuum of the original operators $c_{\mathbf{k}\sigma }$ with
down spin electrons. As usual in the grand canonical ensemble, the ground
state is then obtained by filling the vacuum of $\psi _{\mathbf{k}}$ with
all the negative energy quasiparticles i.e. with $\gamma _{2\mathbf{k}%
}^{\dagger }$ and $\gamma _{3\mathbf{k}}^{\dagger }$ whose energies, $E_{%
\mathbf{k}}^{2}=-E_{+}(\mathbf{k})$ and $E_{\mathbf{k}}^{3}=-E_{-}(\mathbf{k}%
)$ respectively, are negative for all $\mathbf{k}$. One recovers the BCS and
SDW ground states in the appropriate limits. In general that ground state
contains triplet pairs.

\subsection{Combined spin Josephson effect and ordinary Josephson effect in
superconducting antiferromagnets}

We now turn to the calculation of the spin and charge tunneling current
across the junction. In terms of the $\gamma $'s, the equation for $d\mathbf{%
\hat{S}}_{L}/dt$ (\ref{Kubo1}) can be rewritten as 
\begin{widetext}
\begin{equation}\label{Rappel_Kubo1_transfo}
\begin{aligned}
	\dot{\mathbf{S}}_{G}(t) &=\dfrac{1}{N}\sum^{*}_{\mathbf{k},\mathbf{q}}
	\sum_{\beta}\sum_{i,j}\text{Im}\Big[\exp{i\Delta\phi/2} t_{\mathbf{k}\mathbf{q}}\,\Big\{\vec{%
\sigma }_{ \uparrow\beta }%
	\mathbf{U}_{\beta \uparrow } \tilde{\Gamma}%
^{ij}_{\mathbf{kq}\uparrow\uparrow}\big\langle \gamma^{\dag 
	}_{i\mathbf{k}%
}(t)\gamma_{j\mathbf{q}}(t)\big\rangle 
	+\vec{\sigma }_{ \uparrow\beta
}%
	\mathbf{U}_{\beta \downarrow } \tilde{\Gamma}^{ij}_{\mathbf{kq}%
\uparrow\downarrow}\big\langle \gamma^{\dag 
	}_{i\mathbf{k}%
}(t)\gamma^{\dag}_{j\mathbf{q}}(t)\big\rangle \\
	&\qquad\qquad\qquad+\vec{%
\sigma }_{\downarrow\beta }%
	\mathbf{U}_{\beta \uparrow } \tilde{\Gamma}%
^{ij}_{\mathbf{kq}\downarrow\uparrow}\big\langle \gamma_{i\mathbf{k}%
}(t)
	\gamma_{j\mathbf{q}}(t)\big\rangle 
	+\vec{\sigma }_{
\downarrow\beta }%
	\mathbf{U}_{\beta \downarrow } \tilde{\Gamma}^{ij}_{%
\mathbf{kq}\downarrow\downarrow}\big\langle \gamma_{i\mathbf{k}%
}(t)
	\gamma^{\dag}_{j\mathbf{q}}(t)\big\rangle 
	\Big\}	
	\Big]\;,
\end{aligned}
\end{equation}
\end{widetext}where the phase difference and the difference in quantization
axis have been gauged into the tunneling matrix element and where we defined 
\begin{equation}
\begin{aligned}
\tilde{\Gamma}^{ij}_{\mathbf{kq}\uparrow\uparrow}&=[A_{1i}(%
\mathbf{k})+A_{3i}(\mathbf{k})][A_{1j}(\mathbf{q})+A_{3j}(\mathbf{q})]\;,\\
\tilde{\Gamma}^{ij}_{\mathbf{kq}\uparrow\downarrow}&=[A_{1i}(%
\mathbf{k})+A_{3i}(\mathbf{k})][A_{2j}(\mathbf{q})+A_{4j}(\mathbf{q})]\;,\\
\tilde{\Gamma}^{ij}_{\mathbf{kq}\downarrow\uparrow}&=[A_{2i}(%
\mathbf{k})+A_{4i}(\mathbf{k})][A_{1j}(\mathbf{q})+A_{3j}(\mathbf{q})]\;,\\
\tilde{\Gamma}^{ij}_{\mathbf{kq}\downarrow\downarrow}&=[A_{2i}(%
\mathbf{k})+A_{4i}(\mathbf{k})][A_{2j}(\mathbf{q})+A_{4j}(\mathbf{q})]\;.
\end{aligned}  \label{gamma_tilde}
\end{equation}%
We compute each average in (\ref{Rappel_Kubo1_transfo}) to first order in
the tunneling Hamiltonian as in Eq.(\ref{Kubo2}) which requires $H_{T}$ to
be also rewritten in terms of the $\gamma $'s. However, unlike the previous
case of itinerant antiferromagnetism, all the terms in $H_{T}$ must now be
included since unconventional pairing is allowed by the presence of
superconductivity. There will appear two-point correlation functions of the
form $\langle c_{\mathbf{k}\sigma }^{\dag }(t)c_{-\mathbf{k}-\sigma }^{\dag
}(t^{\prime })d_{\mathbf{q}\sigma }(t)d_{-\mathbf{q}-\sigma }(t^{\prime
})\rangle _{0}$, corresponding to the coherent tunneling of Cooper pairs,
and these will acquire a factor $\exp (i\Delta \varphi )$, where $\Delta
\varphi $ is the phase difference between the superconducting order
parameters of the two leads. Taking this into account, the calculation of
the staggered magnetic moment current leads to 
\begin{equation}
\dot{\mathbf{S}}_{G}(t)=\big(I_{c}+J_{c}\cos \Delta \varphi \big)\hat{s}%
_{D}\times \hat{s}_{G}\;,  \label{expressionFinale_courant_spin_coex}
\end{equation}%
where 
\begin{widetext}
\begin{equation}
	I_{c}\equiv \dfrac{1}{N^{2}}\sum^{*}_{\substack{\mathbf{k},\mathbf{q}\\ i,j}}|t_{\mathbf{kq}}|^{2}\,\Bigg[
	\Big[(\tilde{\Gamma}^{ij}_{\mathbf{kq}\uparrow 	
		\uparrow})^{2}+(\tilde{\Gamma}^{ij}_{\mathbf{kq}\downarrow\downarrow})^{2}\Big]
		\text{P}\bigg(\dfrac{f(E^{j}_{\mathbf{q}})-f(E^{i}_{\mathbf{k}})}{E^{j}_{\mathbf{q}}-E^{i}_{\mathbf{k}}}\bigg)
		+\Big[(\tilde{\Gamma}^{ij}_{\mathbf{kq}\uparrow\downarrow})^{2}+(\tilde{\Gamma}^{ij}_{\mathbf{kq}\downarrow\uparrow})^{2}\Big]
	\text{P}\bigg(\dfrac{1-f(E^{j}_{\mathbf{q}})-f(E^{i}_{\mathbf{k}})}{E^{j}_{\mathbf{q}}+E^{i}_{\mathbf{k}}}\bigg)\Bigg]
\end{equation}
\end{widetext}and 
\begin{align}
J_{c}\equiv & \dfrac{1}{N^{2}}\sum_{\substack{ \mathbf{k},\mathbf{q}  \\ i,j
}}^{\ast }|t_{\mathbf{kq}}|^{2}\,\tilde{\Gamma}_{\mathbf{kq}\uparrow
\uparrow }^{ij}\tilde{\Gamma}_{\mathbf{kq}\downarrow \downarrow }^{ij}\,%
\text{P}\Bigg[\dfrac{f(E_{\mathbf{q}}^{j})-f(E_{\mathbf{k}}^{i})}{E_{\mathbf{%
q}}^{j}-E_{\mathbf{k}}^{i}}  \notag \\
& +\dfrac{1-f(E_{\mathbf{q}}^{j})-f(E_{\mathbf{k}}^{i})}{E_{\mathbf{q}%
}^{j}+E_{\mathbf{k}}^{i}}\Bigg]
\end{align}%
with $E_{\mathbf{k}}^{i}$ defined at the end of the previous subsection.

Similarly, we obtain the charge supercurrent (or Josephson current) across
the junction by computing the statistical average of the time derivative of
the number operator on the left lead, $N_{G}=\sum_{\mathbf{k}\sigma }c_{%
\mathbf{k}\sigma }^{\dag }c_{\mathbf{k}\sigma }$: 
\begin{equation}
\dot{N}_{G}(t)=\big(\bar{I}_{c}+\bar{J}_{c}\cos \theta \big)\sin \Delta
\varphi \;,  \label{expr_finale_courant_supra_coex}
\end{equation}%
with 
\begin{align}
\bar{I}_{c}\equiv & \dfrac{2}{N^{2}\hbar }\sum_{\substack{ \mathbf{k},%
\mathbf{q}  \\ i,j}}^{\ast }|t_{\mathbf{kq}}|^{2}\tilde{\Gamma}_{\mathbf{kq}%
\uparrow \uparrow }^{ij}\tilde{\Gamma}_{\mathbf{kq}\downarrow \downarrow
}^{ij}\,\text{P}\Bigg[\dfrac{f(E_{\mathbf{q}}^{j})-f(E_{\mathbf{k}}^{i})}{E_{%
\mathbf{q}}^{j}-E_{\mathbf{k}}^{i}}  \notag \\
& -\dfrac{1-f(E_{\mathbf{q}}^{j})-f(E_{\mathbf{k}}^{i})}{E_{\mathbf{q}%
}^{j}+E_{\mathbf{k}}^{i}}\Bigg]\;,
\end{align}%
and $\bar{J}_{c}$ is obtained from $\bar{I}_{c}$ by changing the sign
between the two quotients in the brackets. Note that $\bar{J}%
_{c}=2J_{c}/\hbar .$We discuss the consequences of this equality in the next
section. 

\section{Discussion and conclusion}

The robustness of the spin-Josephson effect is related to the rigidity of a
broken symmetry, in analogy with the ordinary Josephson effect. From a
phenomenological point of view, the common thread between the generalized
Josephson effects described in this work is that the effective
Ginzburg-Landau free energy, or the effective action, contains interaction
terms proportional to product of the order parameters on the left and on the
right of the tunnel junction. Symmetry imposes $\left( \Psi _{L}\Psi
_{R}^{\ast }+h.c.\right) \propto \cos \left( \Delta \phi \right) $ in the
superconducting case and $\mathbf{S}_{L}\left( \mathbf{Q}\right) \cdot 
\mathbf{S}_{R}\left( -\mathbf{Q}\right) $ in the antiferromagnetic case,
with $\mathbf{S}_{L}\left( \mathbf{Q}\right) $ the N\'{e}el vector. The
latter case is completely analogous to the ferromagnetic one where one finds
the scalar product of the uniform magnetizations \cite{Lee, Nogueira, wang}.
The corresponding equations of motion involve $\sin \left( \Delta \phi
\right) $ in the superconducting case and its natural generalization in the
vector case, namely the cross product. The resulting antiferromagnetic spin
current leads to a precession of the order parameters about their sum on
either side of the junction. In the case of a spin $1/2,$ this precession
can be detected experimentally by comparing the magnetic resonance of the
ferromagnet in an AF/I/FM junction with the magnetic resonance of a
AF/I/AF/I/FM junction, as we have shown. We speculate that similar effects
exist for spin one antiferromagnetic junctions in which case the effect
could be directly observed through antiferromagnetic resonance \cite{Kittel}.

In the mean-field solution, the direction of the triplet in spin space is
locked to that of the antiferromagnetic order parameter. For a $d$ -wave
superconducting antiferromagnet, there is an additional cross term in the
Ginzburg-Landau free energy that comes from the triplet component $%
\overrightarrow{\Psi }^{t}$of the order parameter, namely $\left( 
\overrightarrow{\Psi }_{L}^{t}\cdot \overrightarrow{\Psi }_{R}^{t\ast
}+h.c.\right) $ that leads to a) a modulation of the critical charge current
by the dot product of the antiferromagnetic order parameters and b) a
modulation of the critical spin current by the superconducting phase
difference. An analogous effect was found for superconducting ferromagnets 
\cite{Sudbo:2007}. The equality $\bar{J}_{c}=2J_{c}/\hbar $ found at the end
of the previous section is a direct consequence of the fact that both cross
terms come from a single term in the effective Ginzburg-Landau free energy.
There is thus a reciprocity between the modulation of the Josephson charge
current by the antiferromagnetic order parameter and the modulation of the
antiferromagnetic Josephson spin current by the superconducting order
parameter.

An additional analogy between the ordinary Josephson effect and the
spin-Josephson effect resides in gauge transformations. In the ordinary
Josephson effect, a gauge transformation allows the phase difference to be
reported entirely on the tunneling matrix element. Similarly, in the spin
Josephson effect, one can make a choice of quantization axis (analogous to a
gauge choice) that transfers to the tunneling matrix element everything that
is related to the difference in order parameter orientation across the
junction.

We learned from the microscopic calculation that tunneling of Cooper pairs
is replaced by tunneling of the corresponding condensed objects in the
antiferromagnetic case, namely spin-one, charge zero, particle-hole pairs,
with zero projection along the direction of the antiferromagnetic order
parameter. Alternatively, and also in analogy with the superconducting case,
the terms that lead to the spin-Josephson effect represent interference in
the tunneling process between momentum $\mathbf{k+Q}$ spin up particles and
momentum $-\mathbf{k}$ spin-down hole. In superconducting antiferromagnets,
both types of condensed objects, Cooper pairs and finite $\mathbf{Q}$
particle-hole pairs, are present hence one sees both Josephson and
spin-Josephson effects. In addition, in superconducting antiferromagnets a
pure singlet state between nearest neighbors is impossible because of the
broken spin symmetry between the two sublattices. Hence there are always
triplet Cooper pairs with total wave vector $\mathbf{Q}$\cite{w_egal_0}%
\textbf{. }It is the tunneling of this type of Cooper pair that leads to the
cross term between charge and spin-Josephson effects mentioned above.
Indeed, a triplet Cooper pair carries both phase and spin information and
hence its tunneling can influence both spin and charge and can be
manipulated by fields that couple to either spin or charge.

One of the most far reaching consequences of the Josephson effect in
ordinary superconductors is the AC effect where phase difference between the
superconductors increases linearly with time due to an applied voltage
difference. The corresponding frequency of the current is proportional to
the voltage drop through a material independent universal constant $2e/h.$
In the case of the spin Josephson effect, even without external field there
is a precession frequency. In the presence of a uniform magnetic field, the
two N\'{e}el vectors precess around their sum while that sum precesses
around the magnetic field. The correct analog of the AC effect corresponds
to adding fields that will lead to a time variation of the angle between the
the two order parameters, the analog of the phase difference. Adding a
uniform magnetic field does not achieve this. It is only in the presence of
a magnetic field difference between the two sides of the junction that the
angle becomes time dependent, as described by Eq.(\ref{AC_spin_Josephson}), $%
\dot{\vartheta}(t)=-g\mu _{B}\delta \mathbf{B}\cdot \hat{\mathbf{e}}(t)$
with $\delta \mathbf{B}$ the magnetic field difference and $\hat{\mathbf{e}}%
(t)\equiv \hat{\mathbf{s}}_{R}(t)\times \hat{\mathbf{s}}_{L}(t).$ If the
correct order parameters are used, this result is valid for tunneling
between ferromagnets as well. As far as we know, no analog if the equation $%
\dot{\vartheta}(t)=-g\mu _{B}\delta \mathbf{B}\cdot \hat{\mathbf{e}}(t)$ has
appeared in the literature on spin-Josephson effect. Contrary to the
ordinary Josephson effect, the proportionality constant in front of $\delta 
\mathbf{B}$ is not universal since $g$ is material dependent. This is
because the coupling of the magnetic field to matter is not a pure gauge
coupling. In the presence of spin-orbit interactions, both charge and spin
currents couple to the field.

One possible application of our work is to use it to differentiate between
homogeneous coexistence of antiferromagnetism and d-wave superconductivity
and phase separation between these two types of order. It would probably be
easiest to measure time dependent modulation of the "charge" Josephson
critical current induced by a junction where a magnetic field gradient small
enough not to destroy superconductivity is applied to the tunnel junction.
In the case of phase separation, the cross terms that we described are not
present since they come from tunneling of a finite momentum triplet Cooper
pair that exists only when there is homogeneous coexistence. The issue of
homogeneous coexistence is important for the cuprates, the heavy fermions,
the organics and the pnictides where antiferromagnetism and d-wave
superconductivity often seem to overlap in the phase diagram.

Further possible generalizations of our work include a) Studying tunneling
between states with more complicated order parameters b) Calculating the
analogs of the effects that we found in the case where the tunnel junction
is replaced by a normal metal. It is known \cite%
{Ishii:1970,Ovsyannikov:2009,Brydon:2009} that Andreev states in the
superconducting case profoundly change the dependence of the current on the
phase difference, but not its periodicity. Since the analog of Andreev
states exist in the antiferromagnet \cite{Bobkova:2005}, similar effects are
expected for the spin-Josephson effect.

\begin{acknowledgments}
We are grateful to M.B. Paranjape and R.B. MacKenzie for discussions that
stimulated our interest in this problem and to A.V. Andreev, A. Blais, D.
Bergeron, C. Bourbonnais, R. C\^{o}t\'{e}, P. Fournier and D. S\'{e}n\'{e}%
chal for insightful comments. This work was partially supported by FQRNT (Qu%
\'{e}bec), by the Tier I Canada Research Chair Program (A.-M.S.T.) and CIFAR
(A.-M.S.T.).
\end{acknowledgments}


\begin{thebibliography}{99}
\bibitem{half} We limit ourselves to nearest neighbour hopping to simplify
the expression of the transformation amplitudes, but the main result of this
paper, namely formula (\ref{mainresult1}) for the equilibrium current of the
staggered magnetic moment, is also valid for an extended model.
Dissipationless spin current will occur if U is large enough that there is a
gap to quasi-particle excitations.
\end{thebibliography}
\end{document}